\documentclass[preprint,3p,times]{elsarticle}

\usepackage{MCLabPaper_Minimal} 
 
\theoremstyle{remark}

\journal{Journal of Systems and Software}
\newcommand{\PaperTitle}{Scaling Up Statistical Model Checking of Cyber-Physical Systems via Algorithm Ensemble and Parallel Simulations over \Acs{HPC} Infrastructures}

\Input{00-preamble-acronyms.tex}
\Input{00-preamble-notation.tex}
\Input{00-preamble-style.tex}
\Input{00-preamble-database.tex}

\usepackage{mdframed}
\newcommand{\ArticleDisclaimer}{%
	This article appears in Elsevier Journal of Systems and Software, 2025. 
	DOI: \href{https://doi.org/10.1016/j.jss.2024.112238}{10.1016/j.jss.2024.112238}
}

\newcommand{\BoxArticleDisclaimer}{%
	\begin{mdframed}
	\begin{center}
		\ArticleDisclaimer
	\end{center}
	\end{mdframed}
}
\begin{document}

    \begin{frontmatter}



\title{%
    %
    \PaperTitle
}


\author[1]{Leonardo Picchiami\corref{corr1}}
\ead{picchiami@di.uniroma1.it}

\author[2]{Maxime Parmentier\corref{corr1}}
\ead{maxime.parmentier}

\author[2]{Axel Legay}
\ead{axel.legay}

\author[1]{Toni Mancini}
\ead{tmancini@di.uniroma1.it}

\author[1]{Enrico Tronci}
\ead{tronci@di.uniroma1.it}

\cortext[corr1]{Corresponding author}

\affiliation[1]{organization={Computer Science Department, Sapienza University of Rome},
            addressline={Via Salaria 113}, 
            city={Rome},
            postcode={00198}, 
            country={Italy}}

\affiliation[2]{organization={Computer Science Deparment, Universite Catholique de Louvain},
            addressline={Place Sainte Barbe 2}, 
            city={Louvain-la-Neuve},
            postcode={1348}, 
            state={Brabant Wallon},
            country={Belgium}}

\begin{abstract}
Model-based formal verification of industry-relevant \Acp{CPS} is often a computationally prohibitive task. 
In most cases, the complexity of the models precludes any prospect of symbolic analysis, leaving numerical simulation as the only viable option. 
Unfortunately, exhaustive simulation of a \Ac{CPS} model over the entire set of plausible operational scenarios is rarely possible in practice, and alternative strategies such as \Ac{SMC} must be used instead.

In this article, we show that the number of model simulations (samples) required by \Ac{SMC} techniques to converge can be significantly reduced by considering multiple (an \emph{ensemble} of) \Acp{SA} at once, and that the simulations themselves (by far the most expensive step of the entire workload) can be efficiently sped up by exploiting massively parallel platforms.

With three industry-scale \Ac{CPS} models, we experimentally show that the use of an ensemble of two state-of-the-art \Acp{SA} (\ApproxAlgo and \EBGStop) may require dozens of millions fewer samples when compared to running a single algorithm, with reductions in sample size of up to 78\%.
Furthermore, we show that our implementation, by massively parallelizing system model simulations on a \Acs{HPC} infrastructure, yields speed-ups for the completion time of the verification tasks which are practically linear with respect to the number of computational nodes, thus achieving an efficiency of virtually $100\%$, even on very large platforms.
This makes it possible to complete tasks of model-based \Ac{SMC} verification for complex \Acp{CPS} in a matter of hours or days, whereas a na{\"ive} sequential execution would require from months to many years.
\end{abstract}

%


\maketitle
\BoxArticleDisclaimer

\clearpage 

\begin{highlights}
    \AcronymReset{CPS}
    \AcronymReset{EAA}
    \AcronymReset{SA}
    \AcronymReset{SMC}
    \item Model-based verification of industry-relevant \Acp{CPS} is often computationally prohibitive
    \item Simulation-based \Ac{SMC} approaches in the form of \Acp{SA} are often necessary to handle complex models
    \item Our \Ac{EAA} combines multiple \Acp{SA} to minimize the number of required system simulations
    \item Our implementation massively parallelizes model simulations on \Acs{HPC} platforms, and is $\sim 100\%$ efficient even on large infrastructures
    \item As a combined effect of the above, model-based \Ac{SMC} verification tasks of complex \Acp{CPS} are shown to be completed in hours or a few days, whereas a na{\"ive} sequential execution would require from months to many years.
\end{highlights}


\AcronymReset{CPS}
\AcronymReset{EAA}
\AcronymReset{HPC}
\AcronymReset{SA}
\AcronymReset{SMC}
    \end{frontmatter}

    \SectionInput
        [sec:introduction]
        {Introduction}
        {01-introduction.tex}
    \SectionInput
        [sec:background]
        {Preliminaries and related work}
        {02-background.tex}
    \SectionInput
        [sec:formal-framework]
        {Formal framework}
        {03-formal-framework.tex}
    \SectionInput
        [sec:eaa]
        {\Acf{EAA}}
        {04-eaa.tex}
    \SectionInput
        [sec:hpc-eaa]
        {Parallel scenario evaluation over high-performance computing infrastructures}
        {05-hpc-eaa.tex}
    \SectionInput
       [sec:expres]
        {Experiments}
        {06-experiments.tex}
    \SectionInput
        [sec:conclusions]
        {Conclusions and perspectives}
        {07-conclusion.tex}
    \SectionInput*
        {Acknowledgements}
        {08-acknowledgements.tex}

\BoxArticleDisclaimer

\bibliographystyle{elsarticle-num-names} 
\bibliography{root,bibfile_AT}           

\end{document}